\documentclass[journal]{IEEEtran}
\usepackage{amsmath}
\usepackage{cite}
\usepackage{graphicx}
\usepackage{epstopdf}
\usepackage{amsfonts,amsmath,amssymb}
\usepackage{graphicx}
\usepackage{url}
\usepackage{bm}
\usepackage{bbm}
\usepackage{subfigure}
\usepackage{stfloats}
\usepackage{color}

\usepackage[T1]{fontenc}
\usepackage{cite}

\usepackage{graphicx}
\usepackage{makecell}
\usepackage{amsmath,amsthm,amssymb}

\usepackage{algorithm}
\usepackage{amsmath}
\usepackage{algorithmic}

\ifCLASSOPTIONcompsoc
  \usepackage[caption=false,font=normalsize,labelfont=sf,textfont=sf]{subfig}
\else
  \usepackage[caption=false,font=footnotesize]{subfig}
\fi

\usepackage{stfloats}

\fnbelowfloat

\begin{document}
\title{Energy-Efficient Wireless Powered Secure Transmission with Cooperative Jamming for Public Transportation}
\author{Linqing~Gui, ~Feifei~Bao,~Xiaobo~Zhou, ~Chunhua~Yu, ~Feng~Shu, and ~Jun~Li  
\thanks{This work is supported in part by the National Natural Science Foundation of China (Nos. 61602245, 61771244, 61501238, 61702258, and 61472190), in part by the Natural Science Foundation of Jiangsu Province (No. BK20150791), in part by the Open Research Fund of National Key Laboratory of Electromagnetic Environment, China Research Institute of Radiowave Propagation (No. 201500013).}
\thanks{L. Gui, F. Bao, X. Zhou, F. Shu and J. Li are with the Department of Electronic and Optical Engineering, Nanjing University of Science and Technology, Nanjing 210094, China. E-mail: guilinqing@163.com, 1217286069@qq.com, zxb@njust.edu.cn, shufeng0101@163.com, jun.li@njust.edu.cn.}
\thanks{C. Yu is with the School of Electronic Science and Engineering, Nanjing University, Nanjing 210000, China. E-mail: yuchunhua721@163.com.}
}

\maketitle
\begin{abstract}
In this paper, wireless power transfer and cooperative jamming (CJ) are combined to enhance physical security in public transportation networks. First, a new secure system model with both fixed and mobile jammers is proposed to guarantee secrecy in the worst-case scenario. All jammers are endowed with energy harvesting (EH) capability. Following this, two CJ based schemes, namely B-CJ-SRM and B-CJ-TPM, are proposed, where SRM and TPM are short for secrecy rate maximization and transmit power minimization, respectively. They respectively maximize the secrecy rate (SR) with transmit power constraint and minimize the transmit power of the BS with SR constraint, by optimizing beamforming vector and artificial noise covariance matrix. To further reduce the complexity of our proposed optimal schemes, their low-complexity (LC) versions, called LC-B-CJ-SRM and LC-B-CJ-TPM are developed. Simulation results show that our proposed schemes, B-CJ-SRM and B-CJ-TPM, achieve significant SR performance improvement over existing zero-forcing and QoSD methods. Additionally, the SR performance of the proposed LC schemes are close to those of their original versions.

\end{abstract}


%
\IEEEpeerreviewmaketitle

\section{Introduction}
For the sake of green communication, wireless devices are urged to transmit with a very low power. However, due to the broadcast nature of wireless signals, wireless information is still vulnerable to eavesdroppers. Consequently, energy-efficient secure communication has arisen to be an important problem in wireless networks \cite{RHu,RShuArt}. In public places, since eavesdroppers can easily hide themselves and are hard to be distinguished, secure information is easy to be overheard by the eavesdroppers and secure communication is difficult to be guaranteed. Thus, this paper focuses on secure communication issues in public places, especially in city public transportation vehicles. For example, when a manager takes a city train or light rail for one-hour business trip, he would make the best use of the travel time to fulfill commercial tasks via wireless networks, including e-transaction, classified file transfer and email transmission. Meanwhile, a commercial spy/eavesdropper who is disguised as a passenger in the same carriage can easily capture those wireless signals carrying the sensitive information, which may cause huge loss to the company as well as the individual. Therefore, when potential eavesdroppers are detected, security techniques should be employed immediately to protect information transmission.

To address this issue, apart from the traditional encryption techniques at the application layer, physical-layer (PHY) security techniques dedicate to prevent eavesdroppers from intercepting wireless messages, thus enhancing security from the most bottom layer and from the first beginning. One important performance criterion of PHY security is achievable secrecy rate (SR) which is defined as the difference between the transmission rate of the legitimate channel and that of the wiretap channel \cite{R6}. Here the legitimate channel is the channel between the transmission node and the intended destination, while the wiretap channel is the channel between the transmitter and the eavesdropper. A positive SR can be achieved when the wiretap channel is worse than legitimate channel. But if the SR is downgraded below zero, secure transmission will not be guaranteed and the eavesdropper may successfully capture confidential information. In order to improve the secrecy performance of wireless communication systems, many effective schemes have been proposed such as artificial noise \cite{RShuArt,R61}, directional modulation (DM) \cite{RShuSecure,Shu2017Low} and cooperative jamming (CJ) \cite{R10,R11,R12}. Artificial noise is often generated by the transmission node which is equipped multiple antennas so that the noise can be steered to only degrade the wiretap channel. The DM synthesis is achievable by transmitting confidential messages directly towards the desired receivers \cite{RShuSecure}. However, DM is not so feasible in the scenario of city public transportation because it is technically difficult for a remote Base Station (BS) to generate such narrow beams to directionally distinguish the mobile nodes in the same carriage of a public vehicle.

On the contrary, CJ is preferable in the concerned application scenario because mobile devices carried by passengers in the same vehicle are potentially helpful cooperative nodes. The main idea of CJ is that all cooperative nodes assist the transmitter in the secure transmission by generating artificial noise signals to interfere with the eavesdropper. Although CJ can enhance the SR by taking advantage of user cooperation, the good performance is achieved with the help of cooperative nodes which consume their energy to generate and transmit interference signals. One main obstacle that hinder the employment of CJ-based schemes is that cooperative nodes are usually themselves energy starving, e.g., mobile users in the scenario of this article. Therefore, it is critical to fully compensate the energy consumption of cooperative nodes through energy harvesting techniques.

Wireless energy harvesting is an emerging approach to power the energy-constrained networks and help prolonging the lifetime of wireless nodes \cite{R1,FXiao1}. In recent works, radio frequency (RF) energy harvesting techniques are separated into two main families: simultaneous wireless information and power transfer (SWIPT) and wireless powered tranfer (WPT) \cite{R13}. In SWIPT, the transmitted signals carry both energy and information to contemporaneously achieve information delivery and wireless energy recharging \cite{R15}. In contrast, WPT divides wireless communication into two phases. The power transfer phase first broadcasts energy-containing signals to recharge the energy harvesting wireless nodes; then, these nodes transmit packets by utilizing the harvested energy in the previous phase. From the aspect of complexity, WPT is more suitable than SWIPT in the scenario of this article because the cooperative jammers only need the energy of radio signals from power station and they have no interest in the content of those signals.

Although either cooperative jamming or wireless power transfer has been well studied in literature, it is in recent years that their combination has become an attractive research topic \cite{R111,R112,R113,R114}. The authors in \cite{R111} proposed a hybrid base station (BS) which first transfers power to the source and then executes cooperative jamming while the source transmits the information using the harvested energy. However, in the scenario of public transportation, rather than the hybrid BS, it is more reasonable to deploy a power station inside a vehicle to wireless charge cooperative nodes, considering the long distance between the BS and each cooperative node. The wireless-powered network described in \cite{R112} assumed cooperative nodes to be untrusted. In that scenario, the cooperative nodes also acted as relays, i.e., they need to relay the information from the source. The secure network in \cite{R113} comprises of one source, one jammer and one destination. The SR at the destination is maximized by jointly optimizing the power allocation on each subcarrier at the source and the jammer as well as the time allocation between two time slots. In \cite{R114}, the authors provided an overview on cooperative jamming strategies for wireless powered communication networks. Designed for different application scenarios, those CJ strategies cannot be directly employed in public transportation. To the best of our knowledge, this is the first paper investigating the PHY security issue for public transportation. The main contributions of this paper are summarized as follows:

(1) A cooperative jamming based secure communication model with energy harvesting capability is established for public transportation. In this model, cooperative jamming is fulfilled by both fixed jammers and mobile jammers. The fixed jamming nodes pre-installed in the vehicle can help to guarantee basic secrecy performance in the worst-case scenario with no mobile users in the vehicle. On the contrary, when there are other mobile users in the vehicle, they can act as mobile jammers to greatly interfere with the eavesdropper and to maximize the SR. Since mobile jammers consume their limited energy to transmit the interference signals, energy compensation is provided in the model through energy harvesting.

(2) To obtain the best secrecy and power performance, two CJ based optimal schemes are proposed, namely beamforming-CJ-SR-maximization (B-CJ-SRM) and beamforming-CJ-transmit-power-minimization (B-CJ-TPM). These two schemes are designed to maximize the SR and to minimize the transmit power of the BS, respectively. As to B-CJ-SRM, with the constraint on the maximum signal-to-interference-noise-ratio (SINR) of the eavesdropper, the original optimization problem is first converted into a tractable problem, then into a standard semidefinite programming (SDP) problem by the semidefinite relaxation (SDR) technique. Similarly, the original problem of B-CJ-TPM is also transformed into a SDP which can be easily solved by CVX tools. Simulation results show that our schemes have better secrecy and power performance than some existing schemes such as zero-forcing \cite{R11} and {QoSD} \cite{R20}.

(3) Due to the relatively high complexity of the proposed two optimal schemes, we then design two corresponding low-complexity schemes, namely LC-B-CJ-SRM and LC-B-CJ-TPM. These two schemes both employ concave convex procedure (CCCP) iterative method to obtain sub-optimal solutions of their original optimization. The main ideas of two proposed low-complexity schemes are described as follows. First, the optimization problem is transformed into an equivalent difference of convex (DC) programming. Then the CCCP-based iterative method is employed to solve the DC programming. During each iteration, only a second-order cone programming (SOCP) needs to be solved. Finally the complexity of the two proposed schemes are derived, proved to be much lower than that of their original schemes. Simulation results show that our low-complexity schemes have similar performance to our optimal schemes.

The rest of this paper is organized as follows. In Section II, a CJ based secure communication model with energy harvesting capability is introduced. Section III describes our proposed two CJ based schemes namely B-CJ-SRM and B-CJ-TPM. In Section IV, to reduce the complexity of the proposed schemes, we further design their low-complexity versions namely LC-B-CJ-SRM and LC-B-CJ-TPM. Section V presents simulation results to validate the effectiveness and advantage of the proposed schemes. Finally, Section VI concludes the paper.

\textbf{Notations}: In this paper, the lower-case, boldface lower-case and boldface upper-case letters are used to denote scalars, vectors and matrices, respectively. The transpose, conjugate, conjugate transpose, rank and trace of the matrix $X$ are denoted as $X^{T}$, $X^{\ast}$, $X^{H}$, $\text{rank}(X)$ and Tr($X$), respectively. $X\succeq0$ denotes that $X$ is Hermitian positive semidefinite matrix. $E\{\cdot\}$ denotes expectation. $\mathcal{CN}\{\mu,\sigma^{2}\}$ denotes the circularly symmetric, complex Gaussian distribution with mean $\mu$ and variance $\sigma^{2}$. $\log\{\cdot\}$ denotes the base-2 logarithm.

\section{System model}

As shown in Fig. 1, we consider a downlink secure communication system with one Base Station (BS), one destination user, one eavesdropper, one power station and, totally $N$ cooperative nodes. Except the BS, all other nodes are deployed in a public transportation vehicle. The destination user can be actually a mobile user, who is a passenger in the vehicle. The BS is equipped with $M$ antennas and each of all other users is equipped with a single antenna.
The power station is pre-installed in the vehicle for wirelessly transferring power to the cooperative nodes. The cooperative nodes marked as $C_1, C_2, C_3,\ldots,C_N$ are used to transmit jamming signals to deliberately confuse the eavesdropper.
We assume that, among the total $N$ cooperative nodes, the first two nodes are fixed and pre-installed in the vehicle, while the remaining nodes are mobile jamming nodes, which are actually mobile users (e.g., passengers). This assumption is to guarantee a certain level of secrecy in the worst-case scenario, where there is no mobile jamming node in the vehicle, which may happen during the non-peak hours. In this worst-case scenario, the fixed jamming nodes can still transmit jamming signals to create interference at the eavesdropper. Instead of only one node, at least two single-antenna jamming nodes are required in order to enable the jamming nodes to deliberately create different amounts of interference at the eavesdropper and destination user.

\begin{figure}[!h]
\centering
\includegraphics[width=0.45\textwidth]{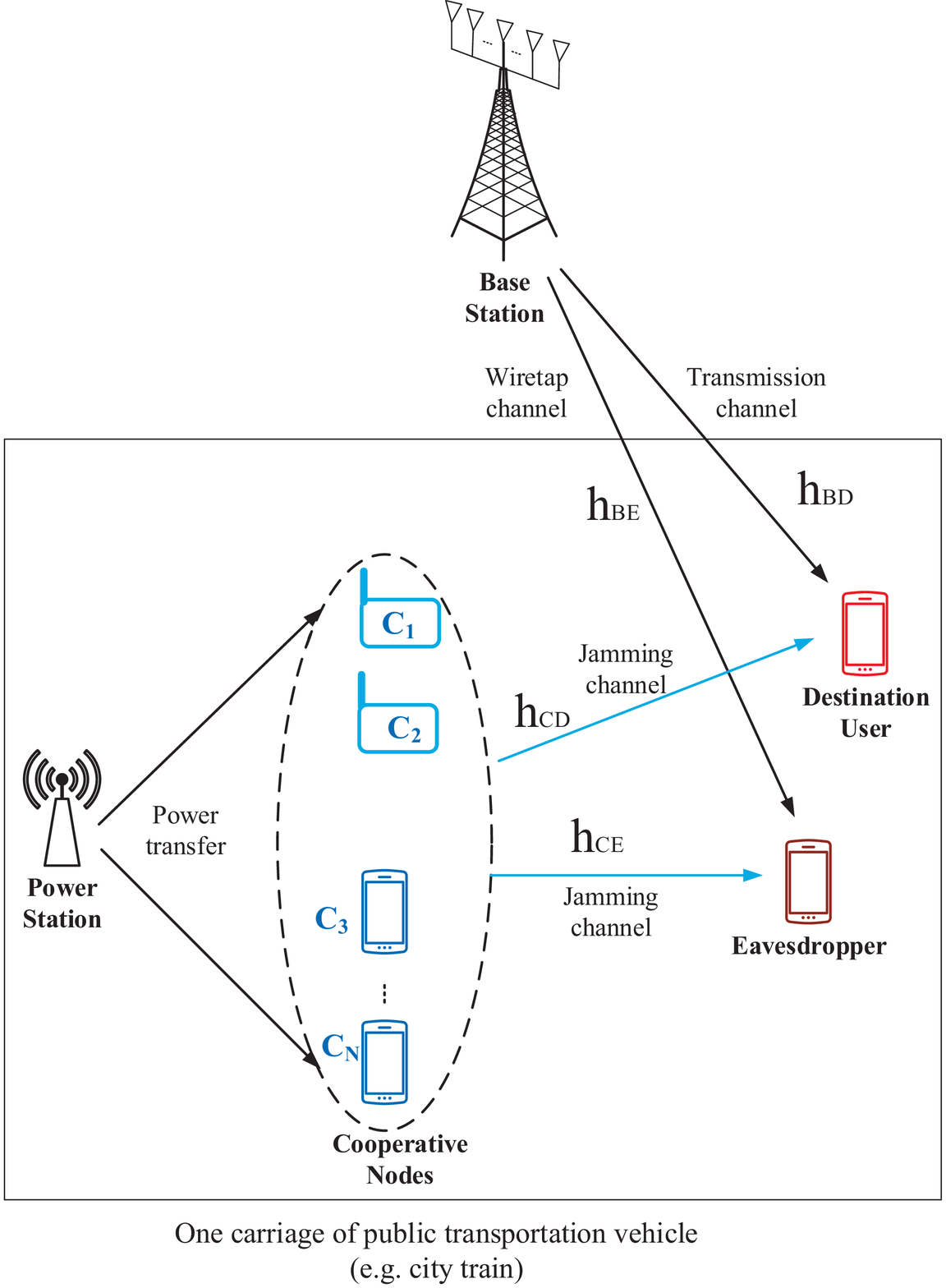}
\caption{System model}
\label{fig1}
\end{figure}

Besides those two fixed cooperative nodes, the mobile users who are also passengers in the vehicle can participate as mobile cooperative jamming nodes. These mobile users are potentially excellent jamming helpers because they may locate close to the eavesdropper. For example, in peak hours, a number of mobile users in the same vehicle can be used to greatly interfere with the eavesdropper such that to improve the secrecy performance. We note that, when these mobile users perform cooperative jamming, they consume the limited energy of their batteries. Compensation or incentive mechanism should be introduced for these helpers. As such, in this work we consider that a power station is available to transfer power to the cooperative jamming nodes. In addition, we assume that when each cooperative node transmits jamming signal, its transmit power should not exceed the power received from the power station.

The information signal vector transmitted from the BS and the jamming signal vector transmitted from the cooperative nodes are denoted by $\mathbf{v}x \in \mathbb{C}^{M \times 1}$ and $\mathbf{z}\in \mathbb{C}^{N \times 1}$ respectively, where $\mathbf{v}$ is beamforming vector adopted by the BS, $x$ represents the confidential information signal for the destination user with $\mathbb{E}[x^{H}x]=1$. Then the received signals at the destination user and the eavesdropper can be expressed as
\begin{align}\label{y_d}
y_d=\mathbf{h}_{BD}^{H}\mathbf{v}x+\mathbf{h}_{CD}^{H}\mathbf{z}+n_d,
\end{align}
and
\begin{equation}\label{y_e}
y_e=\mathbf{h}_{BE}^{H}\mathbf{v}x+\mathbf{h}_{CE}^{H}\mathbf{z}+n_e,
\end{equation}
respectively, where the vectors $\mathbf{h_{BD}} \in \mathbb{C}^{M \times 1}$ and $\mathbf{h_{BE}} \in \mathbb{C}^{M \times 1}$ denote the transmission channel and wiretap channel respectively, while the vectors $\mathbf{h_{CD}} \in \mathbb{C}^{N \times 1}$ and $\mathbf{h_{CE}} \in \mathbb{C}^{N \times 1}$ denote the jamming channels from the $N$ cooperative nodes to the destination user and the eavesdropper, respectively. In this work, we assume that all the channel state information is available for designing the secure system. The assumption that the eavesdropper's channel state information is available can be justified by the fact that the eavesdropper can be a potential legitimate user and thus it has already cooperated with the BS to conduct channel estimation in order to potentially receive information from the BS.
In \eqref{y_d} and \eqref{y_e}, $n_d$ and $n_e$ represent the Additive white Gaussian noise (AWGN) with variance $\sigma^{2}$, at the destination and eavesdropper, respectively, while $\mathbf{z}$ is the zero-mean Gaussian artificial noise (AN) vector with covariance matrix $\mathbf{Q}$, i.e., $\mathbb{E}[\mathbf{z}\mathbf{z}^{H}]=\mathbf{Q}$ and $\mathbf{Q}\succeq0$.

We denote the maximum transmit power of the BS as $P_{BS}$ and then we have
\begin{equation}\label{TP_BS}
\mathbb{E}(|\mathbf{v}x|^{2})=\mathbf{v}^{2}\leq P_{BS}.
\end{equation}
As assumed, the transmit power of each cooperative node is no more than its harvested power from the power station, i.e.,
\begin{equation}\label{TP_CN}
\mathbf{e}_i^{T}\mathbf{Q}\mathbf{e}_i\leq P_i, ~~~i=1,2,...,N.
\end{equation}
where $P_i$ is the harvested power of $i$-th cooperative node, while $\mathbf{e}_i$ is a column vector in which the $i$-th element is 1 and all other elements are all 0's.
Following \eqref{y_d}, the achievable rate from the BS to the destination is given by
\begin{equation}\label{C_d}
R_d=\log\left(1+\frac{|{\mathbf{h}_{BD}}^{H}\mathbf{v}|^2}{{\mathbf{h}_{CD}}^{H}\mathbf{Q}\mathbf{h}_{CD}+\sigma^2}\right).
\end{equation}
Likewise, following \eqref{y_e} the achievable rate from the BS to the eavesdropper is given by
\begin{equation}\label{C_e}
R_e=\log\left(1+\frac{|{\mathbf{h}_{BE}}^{H}\mathbf{v}|^2}{{\mathbf{h}_{CE}}^{H}\mathbf{Q}\mathbf{h}_{CE}+\sigma^2}\right).
\end{equation}

Then, the achievable SR is given by $R_s = \text{max}\{0, R_d-R_e\}$. In this work, a positive SR can be guaranteed due to the known CSI in the considered system model. For example, the BS can transmit the confidential information in the null space of the eavesdropper's channel to guarantee $R_e = 0$ and $R_d >0$ such that $R_s >0$. As such, in this work the achievable SR can be directly written as $R_s = R_d-R_e$. As such, following \eqref{C_d} and \eqref{C_e} the achievable SR as a function of $\mathbf{v}$ and $\mathbf{Q}$ is given by
\begin{align}\label{R_s}
R_s(\mathbf{v},\mathbf{Q})&=\log\left(1+\frac{|{\mathbf{h}_{BD}}^{H}\mathbf{v}|^2}{{\mathbf{h}_{CD}}^{H}\mathbf{Q}\mathbf{h}_{CD}+\sigma^2}\right)\notag\\
&~~-\log\left(1+\frac{|{\mathbf{h}_{BE}}^{H}\mathbf{v}|^2}{{\mathbf{h}_{CE}}^{H}\mathbf{Q}\mathbf{h}_{CE}+\sigma^2}\right).
\end{align}

With the aid of the cooperative nodes, the ultimate goal of the BS is to maximize $R_s(\mathbf{v},\mathbf{Q})$ subject to the power constraints at the BS and the cooperative nodes or to minimize some power consumption while guaranteeing a certain level of SR.
In the following section, we will tackle the optimization of $\mathbf{v}$ and $\mathbf{Q}$ in order to achieve these ultimate goals.

\section{Proposed Joint Design of Secure Beamforming and Cooperative Jamming}

In this section, we joint design the beamforming vector at the BS (i.e., $\mathbf{v}$) and the covariance matrix of the transmitted AN at the cooperative nodes (i.e., $\mathbf{Q}$) in order to achieve different goals of the BS. Specifically, we first propose the B-CJ-SRM scheme to maximize the achievable SR $R_s(\mathbf{v},\mathbf{Q})$ subject to the power constraints at the BS and the cooperative nodes. In addition, we propose the B-CJ-TPM scheme to minimize the power consumption at the BS while guaranteeing $R_s(\mathbf{v},\mathbf{Q}) \geq R_s^0$, where $R_s^0$ is the minimum required value of $R_s(\mathbf{v},\mathbf{Q})$.

\subsection{Proposed B-CJ-SRM}

Originally, our objective is to maximize the SR $R_s(\mathbf{v},\mathbf{Q})$ given in \eqref{R_s} subject to the power constraints at the BS and the cooperative nodes. However, as per \eqref{R_s} we can see that maximizing the achievable SR is to maximize a product of two correlated and generalized eigenvectors, which is a challenging problem to solve. Although a linear search method is employed to solve this kind of problem in \cite{RNegi}, the entire computation is quite complex, resulting in considerable energy consumption. Nevertheless, a less complex solution is feasible if the secure system has a requirement on the maximum SINR of the eavesdropper. To achieve a certain level of secrecy, it is rational to demand the SINR of the eavesdropper stay below a certain value denoted as $\gamma_{e}$. Then a tractable solution can be achieved by reforming the maximization of $R_s(\mathbf{v},\mathbf{Q})$ into the maximization of the destination's SINR. So the optimization problem is given by
\begin{equation}\label{SRM}
\begin{aligned}
  \underset{\mathbf{v,Q}}{\mathop {\max } }\quad&\text{    }\frac{|\mathbf{h}_{BD}^{H}\mathbf{v}{{|}^{2}}}{\mathbf{h}_{CD}^{H}\mathbf{Q}{{\mathbf{h}}_{CD}}+{{\sigma }^{2}}} \\
\text{s.t.}\quad&\text{    }{{\left| \mathbf{v} \right|}^{2}}\le {{P}_{BS}}, \\
 & \text{         }\mathbf{e}_{i}^{T}\mathbf{Q}{{\mathbf{e}}_{i}}\le {{P}_{_{i}}}, ~~i=1,2,...,N, \\
 & \text{         }\frac{|\mathbf{h}_{BE}^{H}\mathbf{v}{{|}^{2}}}{\mathbf{h}_{CE}^{H}\mathbf{Q}{{\mathbf{h}}_{CE}}+{{\sigma }^{2}}}\le {{\gamma }_{e}}, \\
 & \text{          }\mathbf{Q}\succeq \mathbf{0}. \\
\end{aligned}
\end{equation}
Expanding the square terms in \eqref{SRM} and using $\mathbf{V}$ to denote $\mathbf{v}\mathbf{v}^{H}$ (i.e., $\mathbf{V}= \mathbf{v}\mathbf{v}^{H}$), the optimization problem given in \eqref{SRM} can be rewritten as
\begin{equation}\label{SRM_1}
\begin{aligned}
\underset{\mathbf{V},\mathbf{Q}}{\mathop{\max }}\quad&\frac{\mathbf{h}_{BD}^{H}\mathbf{V}{{\mathbf{h}}_{BD}}}{\mathbf{h}_{CD}^{H}\mathbf{Q}{{\mathbf{h}}_{CD}}+{{\sigma }^{2}}} \\
  \text{s.t.}\quad&\text{    Tr}(\mathbf{V})\le {{P}_{BS}}, \\
 & \text{         }\mathbf{e}_{i}^{T}\mathbf{Q}{{\mathbf{e}}_{i}}\le {{P}_{_{i}}},~~i=1,2,...,N, \\
 & \text{         }\frac{\mathbf{h}_{BE}^{H}\mathbf{V}{{\mathbf{h}}_{BE}}}{\mathbf{h}_{CE}^{H}\mathbf{Q}\mathbf{h}_{CE}+{{\sigma }^{2}}}\le {{\gamma }_{e}}, \\
 & \text{          }\mathbf{V}\succeq \mathbf{0},\mathbf{Q}\succeq \mathbf{0}, \\
 & \text{          }\text{rank}(\mathbf{V})=1, \\
\end{aligned}
\end{equation}
We note that $\text{rank}(\mathbf{V})=1$ in \eqref{SRM_1} is a non-convex constraint. For now, we remove this constraint and the optimization problem given in \eqref{SRM_1} is given by
\begin{equation}\label{SRM_2}
\begin{aligned}
\underset{\mathbf{V},\mathbf{Q}}{\mathop{\max }}\quad&\frac{\mathbf{h}_{BD}^{H}\mathbf{V}{{\mathbf{h}}_{BD}}}{\mathbf{h}_{CD}^{H}\mathbf{Q}{{\mathbf{h}}_{CD}}+{{\sigma }^{2}}} \\
 \text{s.t.}\quad&\text{    Tr}(\mathbf{V})\le {{P}_{BS}}, \\
 & \text{         }\mathbf{e}_{i}^{T}\mathbf{Q}{{\mathbf{e}}_{i}}\le {{P}_{_{i}}},~~i=1,2,...,N \\
 & \text{         }\mathbf{h}_{BE}^{H}\mathbf{V}{{\mathbf{h}}_{BE}}-{{\gamma }_{e}}(\mathbf{h}_{CE}^{H}\mathbf{Q}{{\mathbf{h}}_{CE}}+{{\sigma }^{2}})\le 0, \\
 & \text{         }\mathbf{V}\succeq \mathbf{0},\mathbf{Q}\succeq \mathbf{0}. \\
\end{aligned}
\end{equation}
We note that the objective function in \eqref{SRM_2} is quasi-convex, while all other constraints are convex. Fortunately, we can convert the objective function to a convex one by Charnes-Cooper transformation \cite{R19}. Thus the optimization problem given in \eqref{SRM_2} can be again rewritten as
\begin{equation}\label{SRM_3}
\begin{aligned}
 \underset{\mathbf{\tilde{V}},\mathbf{\tilde{Q}},t}{\mathop{\max }}\quad&\text{ Tr}({{\mathbf{H}}_{BD}}\mathbf{\tilde{V}}) \\
 \text{s.t.}\quad&\text{    Tr}(\mathbf{\tilde{V}})\le {t {P}_{BS}}, \\
 & \text{         Tr(}{{\mathbf{e}}_{i}}\mathbf{e}_{i}^{T}\mathbf{\tilde{Q}})\le {t {P}_{_{i}}}\text{ ,  }i=1,2,...,N, \\
 & \text{         Tr}({{\mathbf{H}}_{BE}}\mathbf{\tilde{V}})-{{\gamma }_{e}}(\text{Tr}({{\mathbf{H}}_{CE}}\mathbf{\tilde{Q}})+{{\sigma }^{2}}t)\le 0, \\
 & \text{         Tr}({{\mathbf{H}}_{CD}}\mathbf{\tilde{Q}})+{{\sigma }^{2}}t=1, \\
 & \text{         }\mathbf{\tilde{V}}\succeq \mathbf{0},~~\mathbf{\tilde{Q}}\succeq \mathbf{0}, \\
\end{aligned}
\end{equation}
where $t$ is a slack variable, $\mathbf{\tilde{V}}=t\mathbf{V}$, $\mathbf{\tilde{Q}}=t \mathbf{Q}$, $\mathbf{H}_{BD}=\mathbf{h}_{BD}\mathbf{h}_{BD}^{H}$,   $\mathbf{H}_{CD}=\mathbf{h}_{CD}\mathbf{h}_{CD}^{H}$, $\mathbf{H}_{BE}=\mathbf{h}_{BE}\mathbf{h}_{BE}^{H}$, and $\mathbf{H}_{CE}=\mathbf{h}_{CE}\mathbf{h}_{CE}^{H}$.

Since \eqref{SRM_3} is a standard semidefinite programming (SDP) problem, the optimal solution to it can be found by using SDP solvers such as CVX tools. If the optimal solution of \eqref{SRM_3} is $({{{\mathbf{\tilde{V}}}}^{\ast}}, {{{\mathbf{\tilde{Q}}}}^{\ast}}, t^{\ast})$, then the optimal solution of (\ref{SRM_2}) is $({{\mathbf{V}}^{\ast}}={{{\mathbf{\tilde{V}}}}^{\ast}}/t^{\ast}, {{\mathbf{Q}}^{\ast}}={{{\mathbf{\tilde{Q}}}}^{\ast}}/t^{\ast})$ \cite{RLiOptimal,ROnCJ}. If the rank of ${\mathbf{V}}^{\ast}$ is 1, ${\mathbf{V}}^{\ast}$ can be written as ${{\mathbf{V}}^{\ast}}={{\mathbf{v}}^{\mathbf{*}}}{{\mathbf{v}}^{\mathbf{*}}}^{H}$ based on the eigenvalue decomposition. Therefore, the original optimization problem given in \eqref{SRM} is solved and its optimal solution is $( {\mathbf{v}}^{\mathbf{*}},{\mathbf{Q}}^{\ast})$. We recall that when we transfer the optimization problem \eqref{SRM_1} into \eqref{SRM_2}, the constraint $\text{rank}(\mathbf{V})=1$ is removed. As such, in order to prove that the solution to \eqref{SRM_2} can offer the solution to \eqref{SRM_1}, we only have to prove $\text{rank}({\mathbf{V}}^{\ast}) = 1$. This proof is detailed in Appendix A. Then, the procedure of B-CJ-SRM scheme can be summarized in Algorithm~1.
\begin{algorithm}
\begin{algorithmic}
\STATE \textbf{Input:} $P_{BS}$, $P_i$, $\gamma_e$, $N$, $M$ and $\sigma^{2}$.
\STATE 1. Denote $\mathbf{H}_{BD}$ as $\mathbf{h}_{BD}\mathbf{h}_{BD}^{H}$, $\mathbf{H}_{CD}$ as
$\mathbf{h}_{CD}\mathbf{h}_{CD}^{H}$, $\mathbf{H}_{BE}$ as $\mathbf{h}_{BE}\mathbf{h}_{BE}^{H}$ and $\mathbf{H}_{CE}$ as $\mathbf{h}_{CE}\mathbf{h}_{CE}^{H}$.
\STATE 2. Solve the SDP problem (\ref{SRM_3}) and obtain the optimal solution to \eqref{SRM_3} as $({{{\mathbf{\tilde{V}}}}^{\ast}},{{{\mathbf{\tilde{Q}}}}^{\ast}},t^{\ast})$
\STATE 3. Obtain the optimal solution to (\ref{SRM_2}) as $({{\mathbf{V}}^{\ast}}={{{\mathbf{\tilde{V}}}}^{\ast}}/t^{\ast}, {{\mathbf{Q}}^{\ast}}={{{\mathbf{\tilde{Q}}}}^{\ast}}/t^{\ast})$.
\STATE 4. Obtain $\mathbf{v}^{\ast}$ by performing the eigenvalue decomposition of $\mathbf{V}^{\ast}$.
\STATE \textbf{Output:} $\mathbf{v}^{\ast}$, $\mathbf{Q}^{\ast}$.
\end{algorithmic}
\caption{The Proposed B-CJ-SRM Scheme}\label{algorithm 1}
\end{algorithm}

\subsection{Proposed B-CJ-TPM}
In previous subsection, the SR is maximized with transmit power constraint, i.e., the transmit power of the BS cannot exceed a threshold $P_{BS}$. In that case, to obtain the optimal SR, the actual transmit power of the BS always reaches $P_{BS}$. However, green communication systems are sensitive to energy consumption. Reducing the transmit power of the BS has considerable importance to green communication, because the BS usually consume much more energy than other nodes in the network. To fulfill the coverage, the transmit power of the BS can reach as high as dozens of watts, while the cooperative nodes transmit with much lower power. So how to minimize the transmit power of the BS becomes an important issue. In this subsection, the system is designed with the objective of transmit power minimization with SR constraint, i.e., the SR cannot go below a threshold $R_{s}^{0}$. So the problem of transmit power minimization can be formulated as
\begin{equation}\label{TPM}
\begin{aligned}
\underset{\mathbf{v,Q}}{\mathop{\min \text{ }}}\quad&\text{ }{{\left| \mathbf{v} \right|}^{2}} \\
 \text{s.t.}\quad&\text{  }{{R}_{s}}(\mathbf{v},\mathbf{Q})\ge R_{s}^{0} \\
 & \text{      }\mathbf{e}_{i}^{T}\mathbf{Q}{{\mathbf{e}}_{i}}\le {{P}_{i}}\text{ ,  }i=1,2,...,N \\
 & \text{      }\mathbf{Q}\succeq \mathbf{0}. \\
\end{aligned}
\end{equation}

Similar to the objective function in (\ref{SRM}), the SR constraint in (\ref{TPM}) is also intractable to deal with. To simplify the SR constraint, we replace it with two thresholds which separately limit the destination's and the eavesdropper's SINRs. Then (\ref{TPM}) can be reformulated as
\begin{equation}\label{TPM_1}
\begin{aligned}
\underset{\mathbf{v,Q}}{\mathop{\min \text{ }}}\quad&\text{ }{{\left| \mathbf{v} \right|}^{2}} \\
\text{s.t.}\quad&\text{ } \frac{|\mathbf{h}_{BD}^{H}\mathbf{v}{{|}^{2}}}{\mathbf{h}_{CD}^{H}\mathbf{Q}{{\mathbf{h}}_{CD}}+{{\sigma }^{2}}}\ge {{\gamma }_{d}} \\
 & \text{      }\frac{|\mathbf{h}_{BE}^{H}\mathbf{v}{{|}^{2}}}{\mathbf{h}_{CE}^{H}\mathbf{Q}{{\mathbf{h}}_{CE}}+{{\sigma }^{2}}}\le {{\gamma }_{e}} \\
 & \text{      }\mathbf{e}_{i}^{T}\mathbf{Q}{{\mathbf{e}}_{i}}\le {{P}_{i}}\text{  ,   }i=1,2,...,N \\
 & \text{      }\mathbf{Q}\succeq \mathbf{0}. \\
\end{aligned}
\end{equation}

The relationship between $R_{s}^{0}$ and $(\gamma_d,\gamma_e )$ is $R_{s}^{0}=\log (1+{{\gamma }_{d}})-\log (1+{{\gamma }_{e}})$ .Thus $\gamma_d$ in (\ref{TPM_1}) can be expressed as a function of $R_{s}^{0}$ and $\gamma_e$. Expanding the square terms in (13) and defining $\mathbf{V}$ as $\mathbf{v}\mathbf{v}^{H}$, we can turn the problem (\ref{TPM_1}) to
\begin{equation}\label{TPM_2}
\begin{aligned}
\underset{\mathbf{V,Q}}{\mathop{\min \text{ }}}\quad&\text{  Tr(}\mathbf{V}) \\
\text{s.t.}\quad&\text{  }\frac{\mathbf{h}_{BD}^{H}\mathbf{V}{{\mathbf{h}}_{BD}}}{\mathbf{h}_{CD}^{H}\mathbf{Q}{{\mathbf{h}}_{CD}}+{{\sigma }^{2}}}\ge {{\gamma }_{d}} \\
 & \text{       }\frac{\mathbf{h}_{BE}^{H}\mathbf{V}{{\mathbf{h}}_{BE}}}{\mathbf{h}_{CE}^{H}\mathbf{Q}{{\mathbf{h}}_{CE}}+{{\sigma }^{2}}}\le {{\gamma }_{e}} \\
 & \text{        }\mathbf{e}_{i}^{T}\mathbf{Q}{{\mathbf{e}}_{i}}\le {{P}_{i}}\text{  ,   }i=1,2,...,N \\
 & \text{        }\mathbf{V}\succeq \mathbf{0},\mathbf{Q}\succeq \mathbf{0} \\
 & \text{        }\text{rank}(\mathbf{V})=1. \\
\end{aligned}
\end{equation}

The constraint $\text{rank}(\mathbf{V})=1$ is non-convex, but in fact it can be removed (proof can be found in Appendix B). Without this rank constraint, (\ref{TPM_2}) can be rewritten as
\begin{equation}\label{TPM_3}
\begin{aligned}
\underset{\mathbf{V,Q}}{\mathop{\min }}\quad&\text{  Tr(}\mathbf{V}) \\
 \text{s.t.}\quad&\text{  }{{\gamma }_{d}}(\mathbf{h}_{CD}^{H}\mathbf{Q}{{\mathbf{h}}_{CD}}+{{\sigma }^{2}})-\mathbf{h}_{BD}^{H}\mathbf{V}{{\mathbf{h}}_{BD}}\le 0 \\
 & \text{       }\mathbf{h}_{BE}^{H}\mathbf{V}{{\mathbf{h}}_{BE}}-{{\gamma }_{e}}(\mathbf{h}_{CE}^{H}\mathbf{Q}{{\mathbf{h}}_{CE}}+{{\sigma }^{2}})\le 0 \\
 & \text{       }\mathbf{e}_{i}^{T}\mathbf{Q}{{\mathbf{e}}_{i}}\le {{P}_{i}}\text{  ,   }i=1,2,...,N \\
 & \text{       }\mathbf{V}\succeq \mathbf{0},\mathbf{Q}\succeq \mathbf{0}. \\
\end{aligned}
\end{equation}

Defining $\mathbf{H}_{BD}$ as $\mathbf{h}_{BD}\mathbf{h}_{BD}^{H}$ ,$\mathbf{H}_{CD}$ as $\mathbf{h}_{CD}\mathbf{h}_{CD}^{H}$, $\mathbf{H}_{BE}$ as $\mathbf{h}_{BE}\mathbf{h}_{BE}^{H}$  and $\mathbf{H}_{CD}$ as $\mathbf{h}_{CE}\mathbf{h}_{CE}^{H}$, we can transform (\ref{TPM_3}) into
\begin{equation}\label{TPM_4}
\begin{aligned}
\underset{\mathbf{V,Q}}{\mathop{\min }}\quad&\text{  Tr(}\mathbf{V}) \\
\text{s.t.}\quad&\text{  }{{\gamma }_{d}}(\text{Tr}({{\mathbf{H}}_{CD}}\mathbf{Q})+{{\sigma }^{2}})-\text{Tr(}{{\mathbf{H}}_{BD}}\mathbf{V})\le 0 \\
 & \text{       Tr(}{{\mathbf{H}}_{BE}}\mathbf{V})-{{\gamma }_{e}}(\text{Tr}({{\mathbf{H}}_{CE}}\mathbf{Q})+{{\sigma }^{2}})\le 0 \\
 & \text{       Tr(}{{\mathbf{e}}_{i}}\mathbf{e}_{i}^{T}\mathbf{Q}\text{)}\le {{P}_{i}}\text{  ,   }i=1,2,...,N \\
 & \text{       }\mathbf{V}\succeq \mathbf{0},\mathbf{Q}\succeq \mathbf{0}. \\
\end{aligned}
\end{equation}

Because the problem (\ref{TPM_4}) is a standard SDP problem, its optimal solution denoted as $( {\mathbf{V}}^{\mathbf{*}},{\mathbf{Q}}^{\ast})$ can be found by using SDP solvers such as CVX tools. The rank of ${\mathbf{V}}^{\mathbf{*}}$ is proved to be 1 in Appendix B, thus the rank one constraint in (\ref{TPM_2}) can be removed. Moreover, ${\mathbf{V}}^{\ast}$ can be written as ${{\mathbf{V}}^{\ast}}={{\mathbf{v}}^{\mathbf{*}}}{{\mathbf{v}}^{\mathbf{*}}}^{\mathbf{H}}$  through eigenvalue decomposition. Therefore the problem (\ref{TPM_1}) is solved and its optimal solution is $( {\mathbf{v}}^{\mathbf{*}},{\mathbf{Q}}^{\ast})$.

Finally the procedure of our proposed B-CJ-TPM scheme can be summarized as follows.
\begin{algorithm}
\begin{algorithmic}
\STATE \textbf{Input:} $P_i$, $R_s^{0}$, $\gamma_e$, $N$, $M$ and $\sigma^{2}$.
\STATE 1. Denote $\mathbf{H}_{BD}$ as $\mathbf{h}_{BD}\mathbf{h}_{BD}^{H}$, $\mathbf{H}_{CD}$ as
$\mathbf{h}_{CD}\mathbf{h}_{CD}^{H}$, $\mathbf{H}_{BE}$ as $\mathbf{h}_{BE}\mathbf{h}_{BE}^{H}$ and $\mathbf{H}_{CE}$ as $\mathbf{h}_{CE}\mathbf{h}_{CE}^{H}$.
\STATE 2. Solve the SDP problem (\ref{TPM_4}) and obtain the optimal solution $({{{\mathbf{\tilde{V}}}}^{\ast}},{{{\mathbf{\tilde{Q}}}}^{\ast}})$.
\STATE 3. Obtain $\mathbf{v}^{\ast}$ by performing eigenvalue decomposition for $\mathbf{V}^{\ast}$.
\STATE \textbf{Output:} $\mathbf{v}^{\ast}$, $\mathbf{Q}^{\ast}$.
\end{algorithmic}
\caption{The Proposed B-CJ-TPM Scheme}\label{algorithm 2}
\end{algorithm}

\subsection{Complexity Analysis}

Since solving optimization problems is the major component in all our proposed schemes, the complexity of our schemes depends on the type of the optimization problems and the methods to solve them. In the following, the complexity will be analyzed according to the steps in the literature \cite{R22}.

Our proposed B-CJ-SRM and B-CJ-TPM schemes are finally converted to SDP problems as (\ref{SRM_3}) and (\ref{TPM_4}), respectively. As a result, they can both be solved by CVX software. The solvers used by the CVX software, such as SDPT3, employ a symmetric primal-dual interior-point method. The complexity of this method is derived in \cite{R22} as
\begin{align}\label{complexity_compute}
\mathcal{O}\Big(\big(1+\sum_{j=1}^Jk_j\big)^{\frac{1}{2}}\big(n^3+n^2\sum_{j=1}^Jk_j^2+n\sum_{j=1}^Jk_j^3\big)\log\big(\frac{1}{\epsilon}\big)\Big),
\end{align}
where $\epsilon$ represents a tolerable error or computational accuracy. $k_j$, $J$ and $n$ denote the dimension of the $j$-th constraint, the number of constraints (one equality constraint is equivalent to two inequality constraints), and the total dimensions of all optimization variables, respectively, in an optimization problem.

As for B-CJ-SRM scheme, its SDP optimization problem is (\ref{SRM_3}). Since $J$ equals to the number of constraints in (\ref{SRM_3}), we have $J=N+6$. Similarly, according to the aforementioned definition, we can also derive $k_1=k_2=k_3=...=k_{N+4}=1$, $k_{N+5}=M$, $k_{N+6}=N$, $n=M^{2} + N^{2} + 1$. In order to differentiate the same $n$ for other schemes, we name the parameter $n$ as $n_0$ for B-CJ-SRM, i.e., $n_0=n$. Then, the complexity of B-CJ-SRM is expressed as
\begin{equation}\label{Comp_B-CJ-SRM}
\begin{aligned}
&Comp_{-}B-CJ-SRM(\epsilon)=\\
&\mathcal{O}\Big(\sqrt{M+2N+5}\operatorname{In}\big(\frac{1}{\epsilon }\big)n_0\big(n_0^2+n_0(M^2+N^2+N+4)\\
&+M^3+N^3+N+4\big)\Big).
\end{aligned}
\end{equation}

Similarly, for B-CJ-TPM, since the corresponding SDP optimization problem is (\ref{TPM_4}), we have $J=N+4$, $k_1=k_2=k_3=...=k_{N+2}=1$, $k_{N+3}=M$, $k_{N+4}=N$, $n=n_1= M^2 + N^2$. The complexity of B-CJ-TPM is calculated as
\begin{equation}\label{Comp_B-CJ-TPM}
\begin{aligned}
&Comp_{-}B-CJ-TPM(\epsilon)=\\
&\mathcal{O}\Big(\sqrt{M+2N+3}\operatorname{In}\big(\frac{1}{\epsilon }\big)n_1\big(n_1^2+n_1(M^2+N^2+N+2)\\
&+M^3+N^3+N+2\big)\Big).
\end{aligned}
\end{equation}

Given $M$, then it can be derived from (\ref{Comp_B-CJ-SRM}) and (\ref{Comp_B-CJ-TPM}) that both the complexity of B-CJ-SRM and B-CJ-TPM is approximately $\mathcal{O}(N^{6.5})$. Due to this high complexity, alternative schemes will be investigated in the next section.

\section{Proposed Low-complexity Schemes}

In the scenario of this article, due to the time-varying characteristic of wireless channel, the BS should be able to solve the aforementioned optimization problems as fast as possible, so that the optimal beamforming vector and interference covariance matrix can be renewed in time. Therefore, the proposed schemes should not only keep good secrecy/power performance, but also have low computation complexity. B-CJ-SRM and B-CJ-TPM proposed in last section both have relatively high computation complexity because in order to obtain the optimal performance, rather than directly optimize the beamforming vector $\mathbf{v}$, the two schemes both optimize the matrix $\mathbf{V}=\mathbf{v}\mathbf{v}^{H}$ which has quadratic dimensions compared with $\mathbf{v}$. Therefore, the low complex schemes proposed in this section will directly optimize $\mathbf{v}$. These two schemes namely LC-B-CJ-SRM and LC-B-CJ-TPM both employ concave convex procedure (CCCP) iterative algorithm \cite{R23} to obtain the sub-optimal solutions of the optimization problems in B-CJ-SRM and B-CJ-TPM. The main ideas of two proposed low-complexity schemes are illustrated as follows. Firstly, if necessary, the optimization problem is transformed into an equivalent difference of convex (DC) programming \cite{R24}. Then the CCCP-based iterative algorithm is used to solve the DC programming. During each iteration of the CCCP-based iterative algorithm, only one second-order cone programming (SOCP) \cite{R25} is solved.

\subsection{Proposed LC-B-CJ-SRM}
The optimization problem (\ref{SRM}) is a nonconvex problem because of the nonconvexity of the objective function. Next we first transform problem (\ref{SRM}) into an equivalent DC programming and then solve this DC programming by CCCP-based iterative algorithm. At the first beginning, we rewrite the interference signals transmitted by all cooperative nodes as $\mathbf{q}z$, where $z$ is a random artificial noise with unit power, i.e., $\mathbb{E}(z^{H}z)=1$ and $\mathbf{q}=[q_1,q_2,\ldots,q_N]^{T}$ is a weight vector, in which $q_i$ $(i=1,2,\ldots,N)$ denotes the weight for the $i$th cooperative node. Then we reformulate problem (\ref{SRM}) as
\begin{equation}\label{LSRM}
\begin{aligned}
\underset{\mathbf{v,q}}{\mathop{\max}}\quad&\frac{\mathbf{v}^{H}\mathbf{Av}}{\mathbf{q}^{H}\mathbf{Bq}+\sigma^2}\\
\text{s.t.}\quad&\text{  }\frac{\mathbf{v}^{H}\mathbf{Cv}}{{\mathbf{q}}^{H}\mathbf{Dq}+\sigma^{2}}\le \gamma_e\\
     &\text{  }\mathbf{v}^{H}\mathbf{v}\le P_{BS} \\
     &\text{  }{\mathbf{e}_i}^T\mathbf{q}\mathbf{q}^{H}\mathbf{e}_i\leq P_{i},\,i=1,2,\ldots,N,
\end{aligned}
\end{equation}
where $\mathbf{A}=\mathbf{h}_{BD}\mathbf{h}_{BD}^{H}$, $\mathbf{B}=\mathbf{h}_{CD}\mathbf{h}_{CD}^{H}$, $\mathbf{C}=\mathbf{h}_{BE}\mathbf{h}_{BE}^{H}$ and $\mathbf{D}=\mathbf{h}_{CE}\mathbf{h}_{CE}^{H}$.

By introducing a slack variable $t$, we rewrite (\ref{LSRM}) as
\begin{equation}\label{LSRM_1}
\begin{aligned}
 \underset{\mathbf{v},\mathbf{q},t}{\mathop{\max}}\quad&\text{        }t \\
\text{s.t.}\quad&\text{   }{{\mathbf{q}}^{H}}\mathbf{Bq}+{{\sigma }^{2}}-\frac{{{\mathbf{v}}^{H}}\mathbf{Av}}{t}\le 0 \\
 & \text{       }{{\mathbf{v}}^{H}}\mathbf{Cv}-{{\gamma }_{e}}({{\mathbf{q}}^{H}}\mathbf{Dq}+{{\sigma }^{2}})\le 0\\
 & {{\mathbf{v}}^{H}}\mathbf{v}\le {{P}_{BS}} \\
 & \text{     }{\mathbf{e}_i}^T\mathbf{q}\mathbf{q}^{H}\mathbf{e}_i\leq P_{i},\,i=1,2,\ldots,N \\
 & \text{     }t>0.
\end{aligned}
\end{equation}
It is noted that the functions such as $t$, $\mathbf{v}^{H}\mathbf{Av}$, and ${\mathbf{v}^{H}\mathbf{Av}}/{t}$ ($t>0$, $\mathbf{A}\succeq 0$) are convex \cite{R25}. Hence, problem (\ref{LSRM_1}) is a DC programming. In the following, we will employ the CCCP-based iteration algorithm to find a local optimum of DC programming (\ref{LSRM_1}). Let
\begin{equation}\label{zeta_A}
{\mathbf{\zeta}_\mathbf{A}}(\mathbf{v},t)=\frac{\mathbf{v}^{H}\mathbf{Av}}{t},
\end{equation}
\begin{equation}\label{psi_A}
{\mathbf{\psi}_\mathbf{A}}(\mathbf{v})={\mathbf{v}}^{H}\mathbf{Av}.
\end{equation}
According to the literature \cite{R26}, the first-order Tayor expansions of (\ref{zeta_A}) and (\ref{psi_A}) around the point $(\mathbf{\tilde{v}},\tilde{t})$ are computed as
\begin{equation}\label{zeta_A_Tayor}
{{\zeta }_{\mathbf{A}}}(\mathbf{v},t,\mathbf{\tilde{v}},\tilde{t})=\frac{2\operatorname{Re}\left\{ {{{\mathbf{\tilde{v}}}}^{H}}\mathbf{Av} \right\}}{{\tilde{t}}}-\frac{{{{\mathbf{\tilde{v}}}}^{H}}\mathbf{A\tilde{v}}}{{{{\tilde{t}}}^{2}}}t,
\end{equation}
\begin{equation}\label{psi_A_Tylor}
{{\psi}_{\mathbf{A}}}(\mathbf{v},\mathbf{\tilde{v}})=2\operatorname{Re}\left\{ {{{\mathbf{\tilde{v}}}}^{H}}\mathbf{Av} \right\}-{{{\mathbf{\tilde{v}}}}^{H}}\mathbf{A\tilde{v}}.
\end{equation}
In the $(n + 1)$th iteration of the CCCP-based iterative algorithm, we solve the following convex optimization problem:
\begin{subequations}\label{LSRM_2}
\begin{align}
\underset{\mathbf{v},\mathbf{q},t}{\mathop{\max }}\quad& \text{       }t\\
\text{s.t.}\quad&\text{   }{{\mathbf{q}}^{H}}\mathbf{Bq}+{{\sigma }^{2}}-{{\zeta }_{\mathbf{A}}}(\mathbf{v},t,{{{\mathbf{\tilde{v}}}}^{(n)}},{{{\tilde{t}}}^{(n)}})\le 0 \\
 & \text{       }{{\mathbf{v}}^{H}}\mathbf{Cv}-{{\gamma }_{e}}({{\psi }_{\mathbf{D}}}(\mathbf{q},{{{\mathbf{\tilde{q}}}}^{(n)}})+{{\sigma }^{2}})\le 0 \\
 &\text{       }{{\mathbf{v}}^{H}}\mathbf{v}\le {{P}_{BS}}\\
 & \text{      }{\mathbf{e}_i}^T\mathbf{q}\mathbf{q}^{H}\mathbf{e}_i\leq P_{i},\,i=1,2,\ldots,N.\\
 & \text{        }t>0.
\end{align}
\end{subequations}
where the point $({{\mathbf{\tilde{v}}}^{(n)}}\text{, }{{\mathbf{\tilde{q}}}^{(n)}}\text{, }{{\tilde{t}}^{(n)}})$ denotes the solution to problem (\ref{LSRM_2}) at the $n$th iteration.

We will show that problem (\ref{LSRM_2}) can be further transformed into a SOCP. By letting
\begin{equation}\label{a}
\mathbf{a}=-\frac{2}{{{{\tilde{t}}}^{(n)}}}\mathbf{A}{{{\mathbf{\tilde{v}}}}^{(n)}},
\end{equation}
\begin{equation}\label{b}
b=\frac{{{\left( {{{\mathbf{\tilde{v}}}}^{(n)}} \right)}^{H}}\mathbf{A}{{{\mathbf{\tilde{v}}}}^{(n)}}}{{{\left( {{{\tilde{t}}}^{(n)}} \right)}^{2}}},
\end{equation}

(\ref{LSRM_2}b) is rewritten as
\begin{equation}\label{LSRM_cons}
{{\mathbf{q}}^{H}}\mathbf{Bq}+\operatorname{Re}\left\{ {{\mathbf{a}}^{H}}\mathbf{v} \right\}+bt+{{\sigma }^{2}}\le 0
\end{equation}
which can be converted into a second-order cone constraint, i.e.,
\begin{equation}\label{Cons_SCC}
\left\| \left[ \begin{matrix}
   2{\mathbf{h}_{CD}^{H}}\mathbf{q}  \\
   -\operatorname{Re}\left\{ {{\mathbf{a}}^{H}}\mathbf{v} \right\}-bt-{{\sigma }^{2}}-1  \\
\end{matrix} \right] \right\|\le -\operatorname{Re}\left\{ {{\mathbf{a}}^{H}}\mathbf{v} \right\}-bt-{{\sigma }^{2}}+1.
\end{equation}
Similarly, we can also convert (\ref{LSRM_2}c), (\ref{LSRM_2}d) and (\ref{LSRM_2}e) into second-order cone constraints. Thus, problem (\ref{LSRM_2}) is converted into the following SOCP:
\begin{equation}\label{SOCP_SRM}
\begin{aligned}
&\underset{\mathbf{v},\mathbf{q},t}{\mathop{\max }}\,\text{   }t \\
& \text{s.t.}\\
  &\left\| \left[ \begin{matrix}
   2{\mathbf{h}_{CD}^{H}}\mathbf{q}  \\
   -\!\operatorname{Re}\left\{ {{\mathbf{a}}^{H}}\mathbf{v} \right\}-\!bt-\!{{\sigma }^{2}}-\!1  \\
\end{matrix} \right] \right\|\!\le -\!\operatorname{Re}\left\{ {{\mathbf{a}}^{H}}\mathbf{v} \right\}-\!bt-\!{{\sigma }^{2}}+\!1 \\
 & \left\| \left[ \begin{matrix}
   2{\mathbf{h}_{BE}^{H}}\mathbf{v}  \\
   -\!\operatorname{Re}\left\{{{\mathbf{c}}^{H}}\mathbf{q}\right\}-\!d+{{\sigma}^{2}}-\!{{\gamma }_{e}}  \\
\end{matrix} \right] \right\|\le \!-\!\operatorname{Re}\left\{{{\mathbf{c}}^{H}}\mathbf{q} \right\}-\!d+{{\sigma}^{2}}+\!{{\gamma}_{e}} \\
 & \left\| \mathbf{v} \right\|\le \sqrt{{{P}_{BS}}}\begin{matrix}
   , & \left\| {\mathbf{e}_i}^T\mathbf{q} \right\|\le \sqrt{{{P}_{i}}}, i=1,2,\ldots,N\begin{matrix}
   , & t>0,  \\
\end{matrix}  \\
\end{matrix} \\
\end{aligned}
\end{equation}
where
\begin{equation}\label{c}
\mathbf{c}=-2\mathbf{D}{{{\mathbf{\tilde{q}}}}^{(n)}},
\end{equation}
\begin{equation}\label{d}
d={{\left( {{{\mathbf{\tilde{q}}}}^{(n)}} \right)}^{H}}\mathbf{D}{{{\mathbf{\tilde{q}}}}^{(n)}}.
\end{equation}

Denoting $\theta_1$ as the iteration convergence threshold, we summarize the proposed LC-B-CJ-SRM scheme as Algorithm 3.
\begin{algorithm}
\begin{algorithmic}
\STATE \textbf{Initialization:}
\STATE 1) Given $P_{BS}$, $P_i$, $\gamma_e$, $N$, $M$, $\sigma^{2}$ and $\theta_1$;
\STATE 2) Denote $\mathbf{A}$ as $\mathbf{h}_{BD}\mathbf{h}_{BD}^{H}$, $\mathbf{B}$ as
$\mathbf{h}_{CD}\mathbf{h}_{CD}^{H}$, $\mathbf{C}$ as $\mathbf{h}_{BE}\mathbf{h}_{BE}^{H}$ and $\mathbf{D}$ as $\mathbf{h}_{CE}\mathbf{h}_{CE}^{H}$;
\STATE 3) $n=0$, $(\mathbf{\tilde{v}}^{(n)}, \mathbf{\tilde{q}}^{(n)}, \tilde{t}^{(n)})=(\mathbf{v}_0,\mathbf{q}_0, t_0)$;
\STATE \textbf{Repeat:}
\STATE 1) Solve the problem (\ref{SOCP_SRM}) with $(\mathbf{\tilde{v}}^{(n)}, \mathbf{\tilde{q}}^{(n)}, \tilde{t}^{(n)})$ and obtain the current optimal solution $(\mathbf{v}^{\ast}, \mathbf{q}^{\ast}, t^*)$;
\STATE 2) Update $(\mathbf{\tilde{v}}^{(n+1)}, \mathbf{\tilde{q}}^{(n+1)}, \tilde{t}^{(n+1)})=(\mathbf{v}^{\ast}, \mathbf{q}^{\ast}, t^*)$, $n:=n+1$;
\STATE 3) Compute $|\tilde{t}^{(n+1)}-\tilde{t}^{(n)}|$;
\STATE \textbf{Until:} $|\tilde{t}^{(n+1)}-\tilde{t}^{(n)}|<\theta_1$;
\STATE \textbf{Return:} The local optimal solution of the problem (\ref{LSRM}) $(\mathbf{v}^{\ast}, \mathbf{q}^{\ast})$.
\end{algorithmic}
\caption{The Proposed LC-B-CJ-SRM Scheme}\label{algorithm 3}
\end{algorithm}

\subsection{Proposed LC-B-CJ-TPM}

For the optimization problem (\ref{TPM_1}), we first transform it into an equivalent DC programming
\begin{subequations}\label{LTPM_1}
\begin{align}
\underset{\mathbf{v,q}}{\mathop{\min }}\quad&{{\begin{matrix}
   {} & \left\| \mathbf{v} \right\|  \\
\end{matrix}}^{2}} \\
  \text{s.t.}\quad&\text{     }{{\mathbf{q}}^{H}}\mathbf{Bq}+{{\sigma }^{2}}-\frac{1}{{{\gamma }_{d}}}{{\mathbf{v}}^{H}}\mathbf{Av}\le 0\\
 & \text{         }{{\mathbf{v}}^{H}}\mathbf{Cv}-{{\gamma }_{e}}({{\mathbf{q}}^{H}}\mathbf{Dq}+{{\sigma }^{2}})\le 0\\
 & \text{      }{\mathbf{e}_i}^T\mathbf{q}\mathbf{q}^{H}\mathbf{e}_i\leq P_{i},\,i=1,2,\ldots,N.
\end{align}
\end{subequations}

In the following, the CCCP-based iteration algorithm will be used to solve the DC programming (\ref{LTPM_1}). According to (\ref{psi_A}) and (\ref{psi_A_Tylor}), we derive the $(n + 1)$th iteration of the CCCP-based iterative algorithm to solving the following optimization problem:
\begin{subequations}\label{LTPM_socp}
\begin{align}
\underset{\mathbf{v,q}}{\mathop{\min }}\quad&{{\begin{matrix}
   {} & \left\| \mathbf{v} \right\|  \\
\end{matrix}}^{2}} \\
 \text{s.t.}\quad& \text{    }{{\mathbf{q}}^{H}}\mathbf{Bq}+{{\sigma }^{2}}-\frac{1}{{{\gamma }_{d}}}{{\psi }_{\mathbf{A}}}(\mathbf{v},{{{\mathbf{\tilde{v}}}}^{(n)}})\le 0 \\
 & \text{         }{{\mathbf{v}}^{H}}\mathbf{Cv}-{{\gamma }_{e}}({{\psi }_{\mathbf{D}}}(\mathbf{q},{{{\mathbf{\tilde{q}}}}^{(n)}})+{{\sigma }^{2}})\le 0 \\
 & \text{   }{\mathbf{e}_i}^T\mathbf{q}\mathbf{q}^{H}\mathbf{e}_i\leq P_{i},\,i=1,2,\ldots,N.
\end{align}
\end{subequations}

Similarly, problem (\ref{LTPM_socp}) can also be further converted to a SOCP. Since both (\ref{LTPM_socp}b) and (\ref{LTPM_socp}c) can be written as second-order cone constraint, therefore, problem (\ref{LTPM_socp}) is equivalent to
\begin{equation}\label{SOCP_TPM}
\begin{aligned}
  & \underset{\mathbf{v,q}}{\mathop{\min }}\,{{\begin{matrix}
   {} & \left\| \mathbf{v} \right\|  \\
\end{matrix}}^{2}} \\
 & \text{s.t.}\\
 &\left\| \left[\begin{matrix}
   2{\mathbf{h}_{CD}^{H}}\mathbf{q}  \\
   -\!\frac{1}{{{\gamma }_{d}}}\operatorname{Re}\left\{{{\mathbf{a}}_{1}}^{H}\mathbf{v} \right\}-\!{{b}_{1}}-\!1\\
\end{matrix} \right] \right\|\le-\!\frac{1}{{{\gamma }_{d}}}\operatorname{Re}\left\{ {{\mathbf{a}}_{1}}^{H}\mathbf{v} \right\}-\!{{b}_{1}}+\!1 \\
 &\left\| \left[ \begin{matrix}
   2{\mathbf{h}_{CE}^{H}}\mathbf{v}  \\
   -\!\operatorname{Re}\left\{ {{\mathbf{c}}^{H}}\mathbf{q}\right\}-\!{{d}_{1}}-{{\gamma }_{e}}  \\
\end{matrix} \right]\right\|\le-\!\operatorname{Re}\left\{{{\mathbf{c}}^{H}}\mathbf{q} \right\}-\!{{d}_{1}}+{{\gamma}_{e}} \\
 & \text{        }\left\|{\mathbf{e}_i}^T\mathbf{q} \right\|\le \sqrt{{{P}_{i}}},i=1,2,\ldots,N,\\
\end{aligned}
\end{equation}
where
\begin{equation}
\begin{aligned}
{{\mathbf{a}}_{1}}=-2\mathbf{A}{{{\mathbf{\tilde{v}}}}^{(n)}},
\end{aligned}
\end{equation}
\begin{equation}
\begin{aligned}
{{b}_{1}}=\frac{1}{{{\gamma }_{d}}}{{\left( {{{\mathbf{\tilde{v}}}}^{(n)}} \right)}^{H}}\mathbf{A}{{{\mathbf{\tilde{v}}}}^{(n)}}+{{\sigma }^{2}},
\end{aligned}
\end{equation}
\begin{equation}
\begin{aligned}
{{d}_{1}}={{\left( {{{\mathbf{\tilde{q}}}}^{(n)}} \right)}^{H}}\mathbf{D}{{{\mathbf{\tilde{q}}}}^{(n)}}-{{\sigma }^{2}}.
\end{aligned}
\end{equation}

Denoting $\theta_2$ as the iteration convergence threshold, we summarize the proposed LC-B-CJ-TPM scheme as Algorithm 4.
\begin{algorithm}
\begin{algorithmic}
\STATE \textbf{Initialization:}
\STATE 1) Given $P_i$, $R_s^{0}$, $\gamma_e$, $N$, $M$, $\sigma^{2}$ and $\theta_2$;
\STATE 2) Denote $\mathbf{A}$ as $\mathbf{h}_{BD}\mathbf{h}_{BD}^{H}$, $\mathbf{B}$ as
$\mathbf{h}_{CD}\mathbf{h}_{CD}^{H}$, $\mathbf{C}$ as $\mathbf{h}_{BE}\mathbf{h}_{BE}^{H}$ and $\mathbf{D}$ as $\mathbf{h}_{CE}\mathbf{h}_{CE}^{H}$;
\STATE 3) $n=0$, $(\mathbf{\tilde{v}}^{(n)}, \mathbf{\tilde{q}}^{(n)}, )=(\mathbf{v}_0,\mathbf{q}_0)$;
\STATE \textbf{Repeat:}
\STATE 1) Solve the problem (\ref{SOCP_TPM}) with $(\mathbf{\tilde{v}}^{(n)}, \mathbf{\tilde{q}}^{(n)})$ and obtain the current optimal solution $(\mathbf{v}^{\ast}, \mathbf{q}^{\ast})$;
\STATE 2) Update $(\mathbf{\tilde{v}}^{(n+1)}, \mathbf{\tilde{q}}^{(n+1)}, )=(\mathbf{v}^{\ast}, \mathbf{q}^{\ast})$, $n:=n+1$;
\STATE 3) Compute $|10lg(\|\mathbf{\tilde{v}}^{(n+1)}\|^2)-10lg(\|\mathbf{\tilde{v}}^{(n)}\|^2)|$;
\STATE \textbf{Until:} $|10lg(\|\mathbf{\tilde{v}}^{(n+1)}\|^2)-10lg(\|\mathbf{\tilde{v}}^{(n)}\|^2)|<\theta_2$.
\STATE \textbf{Return:} The local optimal solution of the problem (\ref{LTPM_1}) $(\mathbf{v}^{\ast}, \mathbf{q}^{\ast})$.
\end{algorithmic}
\caption{The Proposed LC-B-CJ-TPM Scheme}\label{algorithm 3}
\end{algorithm}

\subsection{Complexity Analysis}

Similar to the complexity analysis for B-CJ-SRM in Section III, since the LC-B-CJ-SRM focuses on solving SOCP problem (\ref{SOCP_SRM}), its complexity is calculated as follows. In the problem (\ref{SOCP_SRM}), the number of LMI constraints is $J=N+4$, each LMI has a dimension of 1, i.e., $k_1=k_2=k_3=\ldots= k_{N+4}=1$, and the total dimensions of all optimization variables $n=n_2=M+N+1$. According to the formula (\ref{complexity_compute}), the complexity of the LC-B-CJ-SRM is expressed as
\begin{equation}\label{Comp_LC_B-CJ-SRM}
\begin{aligned}
\mathcal{O}\Big(\sqrt{N+5}\operatorname{In}\big(\frac{1}{\epsilon}\big)n_2\big(n_2^2+n_2(N+4)
+N+4\big)\Big)\cdot I_{max},
\end{aligned}
\end{equation}
where $I_{max}$ is the maximum number of iteration times.

Similarly, for LC-B-CJ-TPM, according to its corresponding SOCP optimization problem (\ref{SOCP_TPM}), we have $J=N+2$, $k_1=k_2=k_3=\ldots=k_{N+2} = 1$, $n=n_3=M+N$. The complexity of LC-B-CJ-TPM is calculated as
\begin{equation}\label{Comp_LC_B-CJ-TPM}
\begin{aligned}
\mathcal{O}\Big(\sqrt{N+2}\operatorname{In}\big(\frac{1}{\epsilon}\big)n_3\big(n_3^2+n_2(N+2)
+N+2\big)\Big)\cdot I_{max}.
\end{aligned}
\end{equation}

Given $M$, then it can be derived from (\ref{Comp_LC_B-CJ-SRM}) and (\ref{Comp_LC_B-CJ-TPM}) that both the complexity of LC-B-CJ-SRM and LC-B-CJ-TPM is approximately $\mathcal{O}(N^{3.5})$. Comparing (\ref{Comp_B-CJ-SRM}) with (\ref{Comp_LC_B-CJ-SRM}), we can observe that the LC-B-CJ-SRM has a much lower complexity than the B-CJ-SRM scheme. The same result can be found between LC-B-CJ-TPM and B-CJ-TPM.

\section{SIMULATION AND DISCUSSION}

In this section, we simulate six wireless powered PHY security schemes and evaluate their performance. These six schemes are our proposed two optimal schemes (B-CJ-SRM and B-CJ-TPM), our proposed two low-complexity schemes (LC-B-CJ-SRM and LC-B-CJ-TPM), zero-forcing scheme \cite{R11} and QoSD scheme \cite{R20}. The performance of these schemes will be compared and evaluated in terms of SR as well as transmit power of the BS.

As the simulation scenario, one destination user and one eavesdropper locate in one carriage of city train which drives with a velocity of 60km/h. The distance between the BS and the city train is 400m. The transmit power of the BS ranges from 30dBm to 50dBm. The channel between the BS and the train is assumed to follow Rice distribution, while the channel between any two nodes in the city train is assumed to follow Rayleigh distribution due to the multi-path effect inside city train. The noise is assumed to follow a complex Gaussian distribution with mean 0 and variance $10^{-5}$. Two iteration convergence thresholds $\theta_1$ and $\theta_2$ are 0.01 and 1 respectively.

\begin{figure}[!h]
\centering
\includegraphics[width=0.45\textwidth]{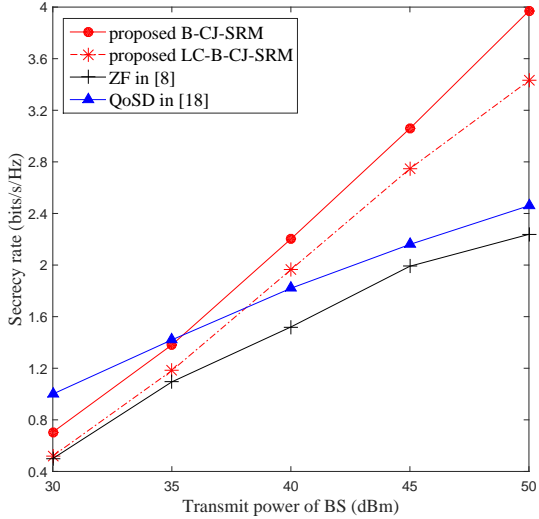}   
\caption{Secrecy rate versus transmit power of the BS, where $M=8$, $N=4$, $P_i=2.5mW$ and $\sigma^{2}=10^{-5}$ }
\label{fig2}
\end{figure}

Fig. 2 shows how the SRs of the aforementioned schemes change with the transmit power of the BS. In the simulation, the BS is equipped with 8 antennas, the number of cooperative nodes is assumed to be 4 (including 2 fixed jamming nodes), while the power harvested by each cooperative node is assumed to be 2.5mW. From this figure, we can observe that when the transmit power of the BS increases, the SR of each scheme also increases. The reason is explained as follows. According to the formula (\ref{R_s}), the SR is determined by the destination's SINR and the eavesdropper's SINR. As the transmit power of the BS increases, both SINRs become larger (note that the power harvested by each cooperative user remains unchanged as well as the power of received interference signals at the destination and eavesdropper). Meanwhile, the destination's SINR increases more than the eavesdropper's SINR, because essentially the optimization problems of the aforementioned schemes all make the efforts to limit the eavesdropper's SINR.

It can also be observed that our proposed low-complexity scheme has similar performance to our proposed B-CJ-SRM scheme. Moreover, our B-CJ-SRM scheme always outperforms the zero-forcing method and in the case of high transmit power (more than 36dBm) also outperforms the QoSD method. The performance gap enlarges with the increase of transmit power. The reason is explained as follows. Our method is to obtain the maximum SINR of the destination, but the zero-forcing method and the QoSD method cannot achieve this, because they only force the interference to the destination to zero or set a minimum threshold for the destination's SINR. These two constraints have limited impact on the growth of the destination's SINR, especially when the BS transmits secure information with a high power. Although the zero-forcing method and the QoSD method aim to minimize the eavesdropper's SINR, compared with the destination's SINR, the eavesdropper's SINR is so small that its reduction can be neglected. From the above anlaysis, we can see that our method has better SR than the other two methods and this advantage becomes wilder when the transmit power increases.

When the transmit power of the BS is small (less than 36dBm), although our method has a slightly higher SR than the zero-forcing method, it performs worse than the QoSD method. The reason is that the QoSD method set a minimum required SINR for the destination in the constraint which guarantees a relatively good SR in case of low transmit power of the BS.

\begin{figure}[!h]
\centering
\includegraphics[width=0.45\textwidth]{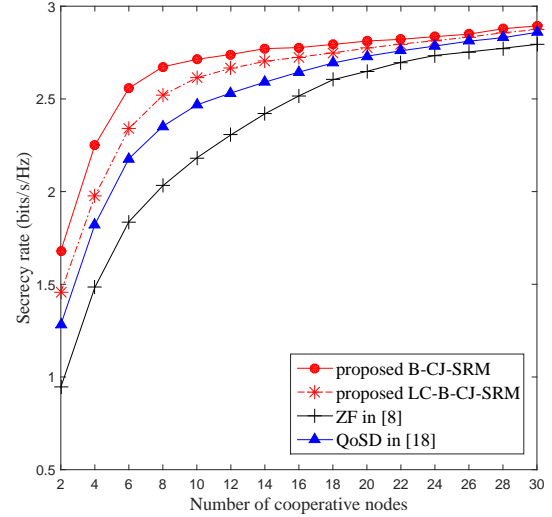}   
\caption{Secrecy rate versus the number of cooperative nodes, where $M=8$, $P_i=2.5mW$, $P_{BS}=10W$}
\label{fig3}
\end{figure}

Fig. 3 shows that the SRs of the aforementioned schemes increase with the number of cooperative nodes. The transmit power of the BS is set to 10W. The reason is explained as follows. With more cooperative nodes, more interference the eavesdropper will suffer. In addition, the transmit power of the BS remains the same. So the eavesdropper's SINR becomes smaller. Meanwhile, the destination's SINR experiences very little or even no degradation because the aforementioned schemes all have restrict requirement on the  destination's SINR. For example, our proposed B-CJ-SRM and its low-complexity version both aim to maximize the destination's SINR, while the QoSD method has a minimum requirement on the destination's SINR. The zero-forcing method directly nulls out the interference to the destination, resulting an unchanging SINR of the destination. Due to a decreasing SINR of the eavesdropper and a relatively stable SINR of the destination, the SRs of the four schemes increase with the number of cooperative nodes.

It can also be observed from Fig. 3, when the number of cooperative nodes exceeds some value (such as 28), the SRs of the four schemes become relativley steady. The reason can be explained as follows. Due to so many cooperative nodes, it is easy to generate great enough interference to the eavesdropper, thus the eavesdropper's SINR will become so small that it approaches to zero. Meanwhile, with so many cooperative nodes, the dimension of interference covariance matrix is high, so it is easy to obtain the optimal interference covariance matrix which can help null out the interference to the destination, resulting in a stable SINR of the destination. Since both SINRs of the destination and the eavesdropper are relatively steady, the SRs of the aforementioned schemes also keep stable when the number of cooperative nodes increases.

\begin{figure}[!h]
\centering
\includegraphics[width=0.45\textwidth]{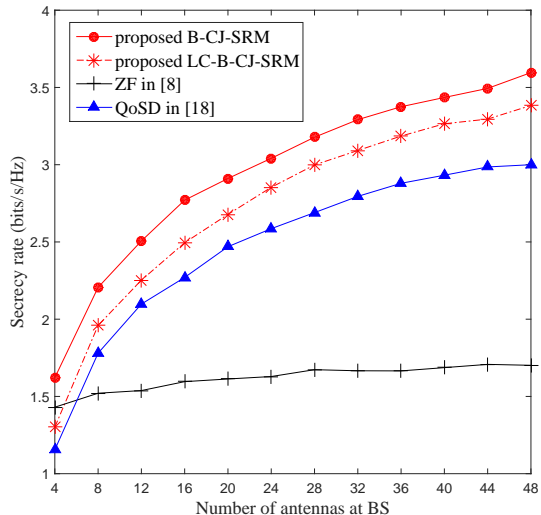}   
\caption{Secrecy rate versus the number of antennas at the BS, where $N=4$, $P_i=2.5mW$, $P_{BS}=10W$}
\label{fig4}
\end{figure}

Fig. 4 shows how the SRs of the concerned schemes change with the number of antennas at the BS. The number of cooperative nodes is set to 4, while the transmit power of the BS is 10W. As the number of antennas at the BS increase, the SRs of our proposed two schemes also increases. The reason is that, by designing the optimal beamforming vector, our schemes can maximize the received power of the secret signal at the desired user as well as the SINR of the desired user. In addition, the SINR of the eavesdropper is limited to a very low value which is much smaller than that of the desired user. When the number of antennas at the BS increases, the beamforming vector has a higher dimension, so it will be easier to design the beamforming vector to maximize the desired user's SINR, then the SR will increase.

From Fig. 4, we can also observe that the SR of the zero-forcing method barely increases with the number of antennas at the BS. The reason is explained as follows. The zero-forcing method only needs to optimize the interference covariance matrix so that the interference to the destination can be nulled out and the interference to the eavesdropper is maximized. Thus there is even no beamforming vector in the optimization problem of the zero-forcing method. Therefore, the increase of the number of antennas at the BS has no significant effect on the performance of the zero-forcing method.

\begin{figure}[!h]
\centering
\includegraphics[width=0.45\textwidth]{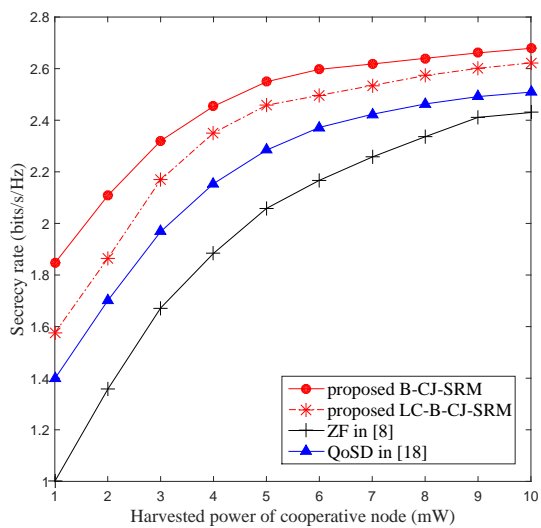}   
\caption{Secrecy rate versus harvested power of cooperative node, where $M=8$, $N=4$, $P_{BS}=10W$}
\label{fig5}
\end{figure}

Fig. 5 shows how the SRs of the four schemes change with the harvested power of each cooperative node. For simplicity, it is assumed that all cooperative nodes received the same power from the power station. The number of cooperative nodes is set to 4, the number of antennas at the BS is fixed as 8, and the transmit power of the BS is 10W. It is observed that as each cooperative node harvests more power from the power station, the SRs of all four schemes get better. The reason is as explained follows. Harvesting more energy, each cooperative node is able to transmit the jamming signal in a higher power, thus increasing the interference to the eavesdropper. Meanwhile, the interference to the destination is effectively limited to a very low level due to the optimization design of the interference covariance matrix. In our proposed schemes, the matrix is well designed to maximize the SINR of the destination, while in zero-forcing scheme the interference to the destination is even nulled out. From all above analysis, it can be concluded that the SRs of the concerned schemes increase with the harvested power of each cooperative node.

From Fig. 5, it can also be observed that when the harvest power of cooperative nodes increases to some extent (more than 8mw for example), the growth of the SRs becomes very slow. The reason is that the relatively high transmit power of interference signal from the cooperative nodes will generate great interference to the eavesdropper, thus effectively reducing the eavesdropper's SINR to nearly 0. In addition, the interference to the destination is limited through the optimization design of the interference covariance matrix. Therefore the SR will become relatively steady.

Fig. 6 shows how the transmit power of the BS changes with the number of cooperative nodes. Here we investigate the transmit power of the BS rather than the transmit power of the cooperative nodes, because the former is much higher (at least 1000 times more) than the latter. Thus the research on the reduction of the transmit power of the BS is much more important than the transmit power of cooperative nodes. The number of antennas of the BS is set to 8, the power harvested by each cooperative node is 2.5mW and the required minimum SR is set to 2bits/s/Hz. From Fig. 6, it can be observed that when there are no more than 12 cooperative nodes, the transmit power of the BS decreases as the number of cooperative nodes increases. The reason is explained as follows. For these four methods, the desired SR is achieved by the collaboration of the BS and the cooperative nodes. To bring forth the same SR, if the cooperative nodes make more contribution, then the BS can contribute less. In the simulation, when the number of cooperative nodes increase, more cooperative nodes will generate greater interference to the eavesdropper, thus having a more positive impact on the desired SR. Meanwhile, in order to keep the same SR, the BS has to reduce its power of secure signal transmission.
\begin{figure}[!h]
\centering
\includegraphics[width=0.45\textwidth]{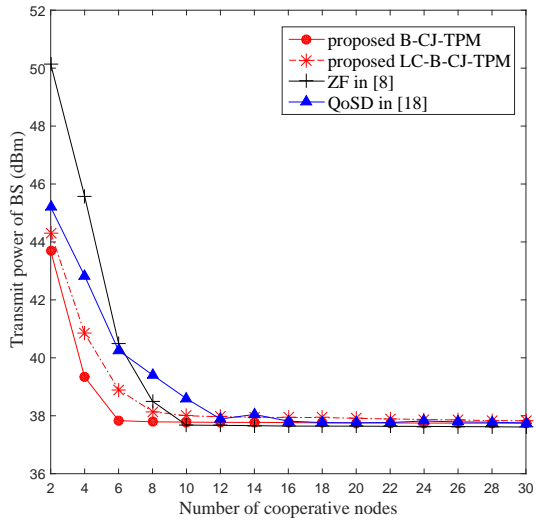}   
\caption{Transmit power of the BS versus the number of cooperative nodes, where $M=8$, $P_i=2.5mW$, required minimum secrecy rate $R_{S}^{0}=2bits/s/Hz$}
\label{fig6}
\end{figure}

It can also be observed from Fig. 6 that when the number of cooperative nodes is relatively large (e.g., greater than 12), the transmit powers of the BS of the four schemes tend to be steady. The reason is that when there are enough cooperative nodes, these nodes will generate considerable interference to the eavesdropper, making the eavesdropper's SINR approach to zero. Meanwhile, with so many cooperative nodes, the dimension of the interference covariance matrix is high, thus it is feasible to obtain the optimal matrix to almost null out the interference to the destination. Then if at the same time the transmit power of the BS is also kept unchanging, the SINR of the destination as well as the SR will become relatively steady. So in order to maintain the SR at 2 bits/s/Hz, the transmit power of the BS will be kept steady.

\section{Conclusion}

This paper investigates wireless powered cooperative jamming technique to enhance energy-efficient security for public transportation. First, a CJ based secure communication model with energy harvesting capability is established. By employing both fixed jammers and mobile jammers to transmit interference signals, the model can not only guarantee basic secrecy performance in the worst-case scenario but also endeavour to greatly interfere with the eavesdropper to obtain the best performance. In addition, to compensate mobile users for their contributions on security, energy compensation is provided in the model through energy harvesting technique. Then to obtain the best secrecy and power performance, we propose two CJ based schemes namely B-CJ-SRM and B-CJ-TPM. The two schemes can maximize the SR (with the transmit power constraint) and minimize the transmit power of the BS (with SR constraint), respectively, by the design of beamforming vector and interference covariance matrix. In order to further reduce the complexity of our proposed optimal schemes, we design a low-complexity versions namely LC-B-CJ-SRM and LC-B-CJ-TPM. Simulation results show that our proposed low-complexity schemes have similar performance to our optimal schemes. Moreover, when the transmit power of the BS is no less than 40dBm, our proposed schemes have significantly better performance than existing zero-forcing and QoSD methods. On the other hand, to achieve the same SR, when there are only a few jammers, our schemes requires less transmit power of the BS than the two existing methods.

\begin{appendices}
\section{}
Since the objective function and constraints in (\ref{SRM_3}) are all convex, the problem (\ref{SRM_3}) meets Slater's condition and Karush-Kuhn-Tucker (KKT) conditions \cite{R21}. The Lagrangian of (\ref{SRM_3}) is
\begin{equation}
\begin{aligned}
& \mathcal{L}(\mathbf{\tilde{V}},\mathbf{\tilde{Q}},t,{{\lambda }_{1}},{{\lambda }_{2}},{{\lambda }_{3}},{{\mu }_{i}}) \\
 & =\text{Tr}({{\mathbf{H}}_{BD}}\mathbf{\tilde{V}})+{{\lambda }_{1}}(\text{Tr}(\mathbf{\tilde{V}})-{{P}_{BS}}t)+ \\
 & \sum\limits_{i=1}^{N}{{{\mu }_{i}}\text{(Tr}({{\mathbf{e}}_{i}}\mathbf{e}_{i}^{T}\mathbf{\tilde{Q}})-{{P}_{i}}t})+{{\lambda }_{2}}(\text{Tr}({\mathbf{H}}_{BE}\mathbf{\tilde{V}})-\\
& \gamma_e(\text{Tr}({{\mathbf{H}}_{CE}}\mathbf{\tilde{Q}})+{{\sigma }^{2}}t))
 +{{\lambda }_{3}}(\text{Tr}({{\mathbf{H}}_{CD}}\mathbf{\tilde{Q}})+{{\sigma }^{2}}t-1)-\\
 &\text{Tr}({{\mathbf{P}}_{1}}\mathbf{\tilde{V}})-
  \text{Tr}({{\mathbf{P}}_{2}}\mathbf{\tilde{Q}}).
\end{aligned}
\end{equation}
where $\lambda_1, \lambda_2, \lambda_3, \mathbf{P}_1, \mathbf{P}_2$ and $\mu_i$($i = 1, 2, \ldots, N$) are all dual variables. The KKT conditions that are relevant to our proof are listed as
\begin{equation}\label{A_1}
{{\left. \frac{\partial L}{\partial \mathbf{\tilde{V}}} \right|}_{\mathbf{\tilde{V}}={{{\mathbf{\tilde{V}}}}^{\ast}}}}={{\mathbf{H}}_{BD}}+{{\lambda }_{1}}\mathbf{I}+{{\lambda}_{2}}{{\mathbf{H}}_{BE}}-{{\mathbf{P}}_{1}}=0
\end{equation}

\begin{equation}\label{A_2}
{{\mathbf{P}}_{1}}{{\mathbf{\tilde{V}}}^{\ast}}=\mathbf{0}
\end{equation}

\begin{equation}\label{A_3}
{{\lambda }_{1}}(\text{Tr}({{\mathbf{\tilde{V}}}^{\ast}})-{{P}_{BS}}{{t}^{\ast}})=0
\end{equation}

\begin{equation}\label{A_4}
{{\lambda }_{1}}\ge 0,{{\lambda }_{2}}\leq0
\end{equation}

\begin{equation}\label{A_5}
{{\mathbf{P}}_{1}}\succeq \mathbf{0},{{\mathbf{\tilde{V}}}^{\ast}}\succeq \mathbf{0}
\end{equation}
where (\ref{A_3}) is a complementary slackness constraint and   is the optimal solution.

Rearrange (\ref{A_1}) and let both sides of it be multiplied by, we can get the following formula
\begin{equation}\label{A_6}
{-\lambda}_{2}{{\mathbf{H}}_{BE}}{{\mathbf{\tilde{V}}}^{\ast}}=({{\lambda }_{1}}\mathbf{I}+{{\mathbf{H}}_{BD}}){{\mathbf{\tilde{V}}}^{\ast}}.
\end{equation}
If $\lambda_1$ can be 0, $\operatorname{Tr}({{{\mathbf{\tilde{V}}}}^{\mathbf{*}}})-{{P}_{BS}}{{t}^{\ast}}$ in (\ref{A_3}) can take any value, which means that the source transmit power (denoted as $\operatorname{Tr}({{{\mathbf{\tilde{V}}}}^{\mathbf{*}}})$) does not definitively equal to the maximum value $P_{BS}$. Thus ${{{\mathbf{\tilde{V}}}}^{\mathbf{*}}}$ in this condition is not the optimal solution. Therefore $\lambda_1$ cannot be equal to 0. Since $\lambda_1>0$, the matrix ${{\lambda }_{1}}\mathbf{I}+{{\mathbf{H}}_{BD}}$ in (\ref{A_6}) is a positive definite matrix and also a full rank matrix. Then we perform the rank operation on both sides of (\ref{A_6}), i.e.,
\begin{equation}
\text{rank}({{\mathbf{H}}_{BE}}{{\mathbf{\tilde{V}}}^{\ast}})=\text{rank}({{\mathbf{\tilde{V}}}^{\ast}}).
\end{equation}

Since $\text{rank}(\mathbf{H}_{BE})=1$, we have
\begin{equation}
\text{rank}({{\mathbf{\tilde{V}}}^{\ast}})=\text{rank}({{\mathbf{H}}_{BE}}{{\mathbf{\tilde{V}}}^{\ast}})\le min(\text{rank}({{\mathbf{H}}_{BE}}),\text{rank}({{\mathbf{\tilde{V}}}^{\ast}})),
\end{equation}
so $\text{rank}({{\mathbf{\tilde{V}}}^{\ast}})\text{ }$ can be 1 or 0. If $\text{rank}({{\mathbf{\tilde{V}}}^{\ast}})\text{ }$ is 0, ${{\mathbf{V}}^{\ast}}$ which equals to
${{\mathbf{v}}^{\mathbf{*}}}{{\mathbf{v}}^{\mathbf{*}}}^{\mathbf{H}}$ becomes an all-zero matrix, thus ${{\mathbf{v}}^{\mathbf{*}}}$ will be an all-zero vector. In this case, there will be no signals transmitted from BS, which is meaningless. Therefore, $\text{rank}({{\mathbf{\tilde{V}}}^{\ast}})\text{ }$ should be 1. Since ${{\mathbf{V}}^{\ast}}$ equals to ${{{\mathbf{\tilde{V}}}}^{\ast}}/t$ and $t^{\ast}$ is a slack variable, we can conclude that $\text{rank}({{\mathbf{V}}^{\ast}})$ equals to 1.
\section{}
The subjective function and constraints in (\ref{TPM_4}) are all convex, so the problem (\ref{TPM_4}) meets Slater's condition and Karush-Kuhn-Tucker (KKT) conditions. The Lagrangian of (\ref{TPM_4}) is
\begin{equation}
\begin{aligned}
  & \mathcal{L}(\mathbf{V},\mathbf{Q},{{\lambda }_{1}},{{\lambda }_{2}},{{\mu }_{r}})= \\
 & \text{Tr(}\mathbf{V})+{{\lambda }_{1}}({{\gamma }_{d}}(\text{Tr}({{\mathbf{H}}_{RD}}\mathbf{Q})+{{\sigma }^{2}})-\text{Tr(}{{\mathbf{H}}_{BD}}\mathbf{V})) \\
 & +{{\lambda }_{2}}(\text{Tr(}{{\mathbf{H}}_{BE}}\mathbf{V})-{{\gamma }_{e}}(\text{Tr}({{\mathbf{H}}_{CE}}\mathbf{Q})+{{\sigma }^{2}})) \\
 & \text{+}\sum\limits_{r=1}^{N}{{{\mu }_{r}}\text{(Tr(}{{\mathbf{e}}_{r}}\mathbf{e}_{r}^{T}\mathbf{Q})-{{P}_{r}}})-\text{Tr}({{\mathbf{P}}_{1}}\mathbf{V})-\text{Tr}({{\mathbf{P}}_{2}}\mathbf{Q}) \\
\end{aligned}
\end{equation}
where $\lambda_1, \lambda_2, \mathbf{P}_1, \mathbf{P}_2$ and $\mu_r$ ($r = 1, 2, \ldots, N$) are all dual variables. We only list the KKT conditions that are relevant to our proof, i.e.,
\begin{equation}\label{B_1}
{{\left. \frac{\partial L}{\partial \mathbf{V}} \right|}_{\mathbf{V}={{\mathbf{V}}^{\ast}}}}=\mathbf{I}-{{\lambda }_{1}}{{\mathbf{H}}_{BD}}+{{\lambda }_{2}}{{\mathbf{H}}_{BE}}-{{\mathbf{P}}_{1}}=0
\end{equation}

\begin{equation}\label{B_2}
{{\mathbf{P}}_{1}}{{\mathbf{V}}^{\ast}}=\mathbf{0}
\end{equation}

\begin{equation}\label{B_3}
{{\mathbf{P}}_{1}}\succeq \mathbf{0},{{\mathbf{V}}^{\ast}}~\succeq \mathbf{0}~
\end{equation}

\begin{equation}\label{B_4}
{{\lambda }_{1}}\ge 0,{{\lambda }_{2}}\ge 0.
\end{equation}

Rearrange (\ref{B_1}) and let both sides of it be multiplied by ${\mathbf{V}}^{\ast}$, we can have the following formula
\begin{equation}\label{B_5}
{{\lambda }_{1}}{{\mathbf{H}}_{BD}}{{\mathbf{V}}^{\ast}}=(\mathbf{I}+{{\lambda }_{2}}{{\mathbf{H}}_{BE}}){{\mathbf{V}}^{\ast}}
\end{equation}
If $\lambda_1$ is 0, the left of (\ref{B_4}) becomes 0, $(\mathbf{I}+{{\lambda }_{2}}{{\mathbf{H}}_{BE}}){{\mathbf{V}}^{\ast}}$ then will also become 0. Since $(\mathbf{I}+{{\lambda }_{2}}{{\mathbf{H}}_{BE}}){{\mathbf{V}}^{\ast}}$ is a full rank matrix, $(\mathbf{I}+{{\lambda }_{2}}{{\mathbf{H}}_{BE}}){{\mathbf{V}}^{\ast}}$ can't be a zero matrix, so
${{\mathbf{V}}^{\ast}}$ has to be 0, which means that there is no signals transmitted from BS. This conflicts with the fact, so $\lambda_1$ cannot be 0. From (\ref{B_4}), we can know that $\lambda_1$ is greater than 0, so $(\mathbf{I}+{{\lambda }_{2}}{{\mathbf{H}}_{BE}}){{\mathbf{V}}^{\ast}}$ in (\ref{B_5}) is a positive definite matrix and also a full rank matrix. Then we perform rank operations on both sides of (\ref{B_5}) and we can have
\begin{equation}
\text{rank}({{\mathbf{H}}_{BD}}{{\mathbf{V}}^{\ast}})=\text{rank}({{\mathbf{V}}^{\ast}}).
\end{equation}

Since $\text{rank}(\mathbf{H}_{BD})=1$, then
\begin{equation}
\text{rank}({{\mathbf{V}}^{\ast}})=\text{rank}({{\mathbf{H}}_{BD}}{{\mathbf{V}}^{\ast}})\le min(\text{rank}({{\mathbf{H}}_{BD}}),\text{rank}({{\mathbf{V}}^{\ast}})).
\end{equation}
If $\text{rank}({{\mathbf{V}}^{\ast}})$ is 0, then ${{\mathbf{V}}^{\ast}}$ is a zero matrix. Since ${{\mathbf{V}}^{\ast}}$ is equivalent to ${{\mathbf{v}}^{\mathbf{*}}}{{\mathbf{v}}^{\mathbf{*}}}^{\mathbf{H}}$, $\mathbf{v}^{\ast}$ will become a zero vector, which conflicts with the fact of signal transmission. Therefore $\text{rank}({{\mathbf{V}}^{\ast}})$ should be 1.

\end{appendices}

\ifCLASSOPTIONcaptionsoff
  \newpage
\fi

\bibliographystyle{IEEEtran}
\bibliography{IEEEabrv,doublecolumn_EH_CJ_PublicTransporation20180501} 
\end{document}